\documentclass[fleqn,twoside]{article}
\usepackage{espcrc2}

\usepackage{epsfig,graphicx}
\usepackage[figuresright]{rotating}

\newcommand{\AmS}{{\protect\the\textfont2
  A\kern-.1667em\lower.5ex\hbox{M}\kern-.125emS}}

\title{ATLAS IBL Pixel Upgrade}
\author{A. La Rosa\address{CERN, CH-1211 Geneve 23, Switzerland}\thanks{Corresponding author. Tel.: +41-(0)76-487-2522. \newline {\it E-mail address}: alessandro.larosa@cern.ch (A. La Rosa)}
             on behalf of ATLAS IBL Collaboration.}
\begin{document}
\begin{abstract}
The upgrade for ATLAS detector will undergo different phase towards super-LHC. The first upgrade for the Pixel detector will consist of the construction of a new pixel layer which will be installed during the first shutdown of the LHC machine (LHC phase-I upgrade).\\ The new detector, called Insertable B-Layer (IBL), will be inserted between the existing pixel detector and a new (smaller radius) beam-pipe at a radius of 3.3 cm. The IBL will require the development of several new technologies to cope with increase of radiation or pixel occupancy and also to improve the physics performance which will be achieved by reducing the pixel size and of the material budget. Three different promising sensor technologies (planar-Si, 3D-Si and diamond) are currently under investigation for the pixel detector.\\
An overview of the project with particular emphasis on pixel module is presented in this paper.
\vspace{1pc}
\end{abstract}
\maketitle
\section{INTRODUCTION}
\begin{figure}[htb]
\vspace{9pt}
\epsfig{figure=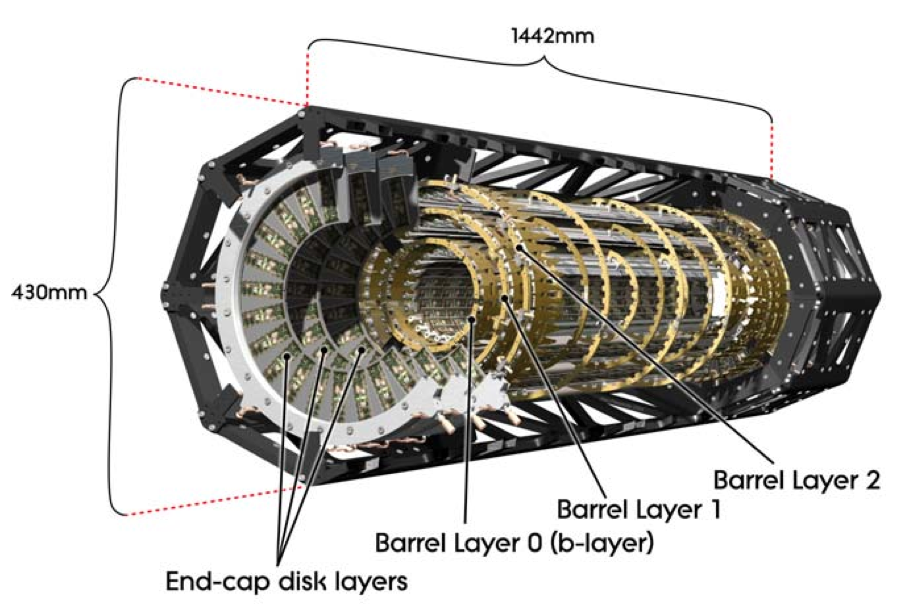, width=0.44\textwidth}
\caption{The current ATLAS Pixel Detector  \cite{PixelDetector}.}
\label{Fig:PixelLayout}
\end{figure}

The ATLAS Pixel Detector \cite{PixelDetector}, shown in Fig.\ref{Fig:PixelLayout} is the innermost part of ATLAS experiment so that it is also the most sensitive sub-detector to radiation damage. Taking into account  the expected lifetime of the Layer-0 (so called B-Layer, which is the closet to the beam pipe) an upgrade of the whole layer is needed after about 5 years of operation, together with the LHC phase-I upgrade.\\ The original program of the B-Layer replacement that foresaw the extraction of the B-Layer and its substitution with a new one has appeared unfeasible. Considering the improvements in many technological aspects and careful engineering of new detector and a smaller beam pipe design, this  allows  to insert a new layer (Insertable B-Layer - IBL) within the existing one.\\
The insertable B-Layer is currently under development and it will be realized using new technologies as new sensors, new front-end chip,  new read-out electronics and new mechanics.\\
The goal of IBL, as 4th pixel layer, is to guarantee the overall performance of the ATLAS \cite{ATLAS} tracking system over the full lifetime of ATLAS. The 4th IBL will be particular important   for the pattern recognition and fake track reduction at high instantaneous luminosity as well as it will provide redundancy in case of failures or radiation damage in the present pixel system.
\section{LAYOUT}
\begin{figure}[htb]
\vspace{9pt}
\epsfig{figure=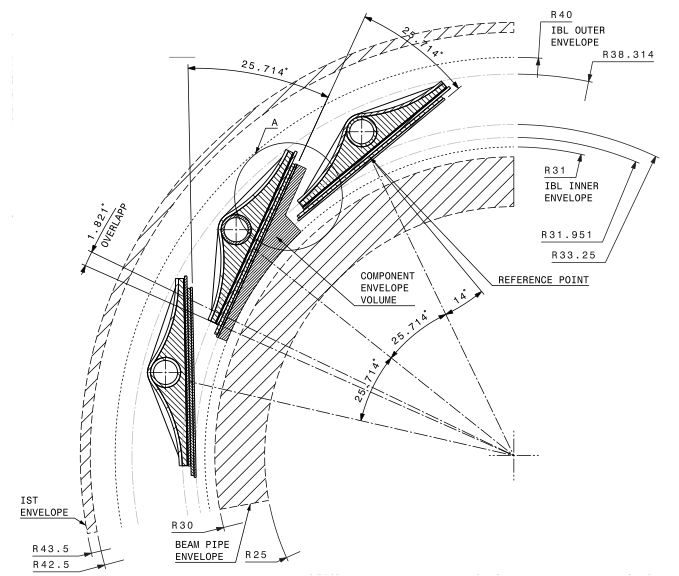, width=0.4\textwidth}
\caption{Baseline of IBL layout: $r$$\phi$ view.}
\label{Fig:IBLLayout}
\end{figure}

A section of the IBL layout is shown in Fig. \ref{Fig:IBLLayout}. It consists of 14 tilted staves (64 cm long and 2 cm wide) with 32 front-end chips per stave and sensors facing the beam pipe. The tilt angle in $\phi$ is almost fixed by space constrains and it is of 14$^\circ$, while the sign is the same as for the Pixel Detector layers \cite{PixelDetector} and has been selected to reduce the Lorentz angle effect of sensors in magnetic field. The inner radius of IBL will be of 31 mm with the outer radius of 38.2 mm while the sensor will present a mean radius of 33 mm.
\section{ELECTRONICS}
The front-end chip foreseen for the IBL is called FE-I4 and it is described in \cite{FEI4}. The integrated circuit has been designed in 130 nm future size bulk CMOS process and it contains readout circuitry for 26880 hybrid pixels organized in a matrix of 80 columns (on 50 $\mu$m pitch) by 336 rows (on 250 $\mu$m pitch).
Each front-end cell contains an independent, free running amplification stage with adjustable shaping, followed by a discriminator with independently adjustable threshold. The FE-I4 keeps track of the firing time of each discriminator as the time over threshold (ToT) with 4-bit resolution, in counts of an externally supplied clock (40 MHz nominal). The chip also keeps tracks of information from all firing discriminators for a latency interval, programmable up to 256 cycles of external clock. Within this latency interval, the information can be traced by supplying a trigger. The data output is serial over a current-balanced pair. The primary output mode is 8b/10b encoded with160 Mb/s rate. The FE-I4 is controlled by a serial LVDS input synchronized by the external clock.\\
The FE-I4 has been submitted at the beginning  of July and it should be available for the first  test in late September 2010.
\section{PIXEL MODULES}
\begin{figure}[htb]
\vspace{9pt}
\epsfig{figure=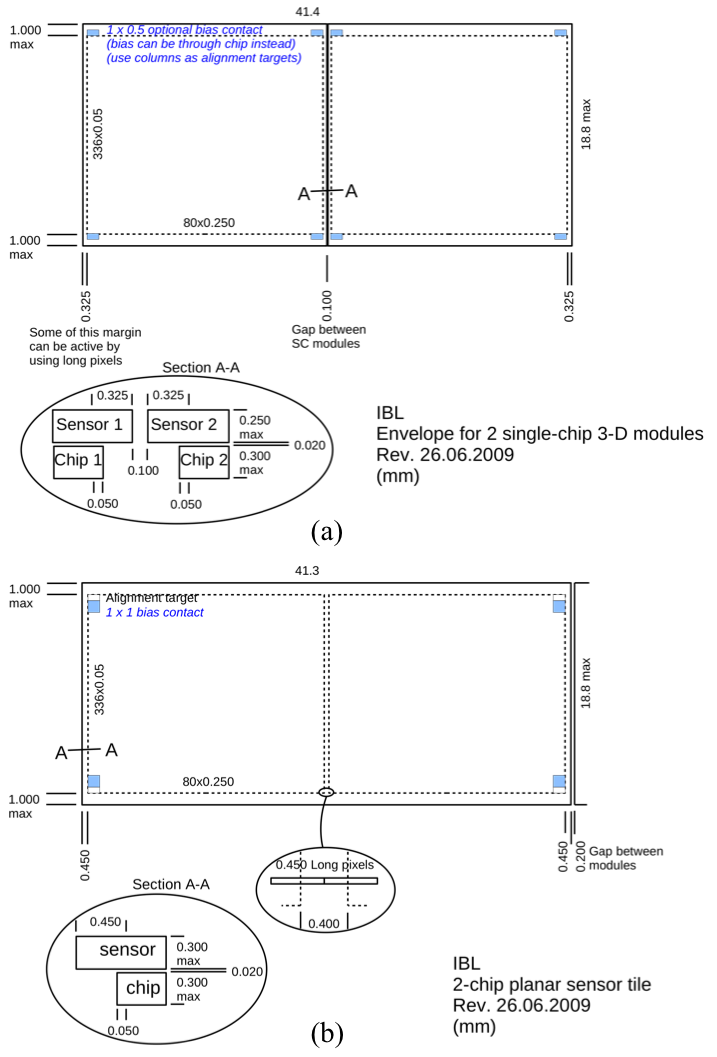, width=0.49\textwidth}
\caption{Module format layout for different sensor types: (a) single-chip module and (b) two-chip module.}
\label{Fig:ModuleLayout}
\end{figure}

A common sensors baseline for engineering and system purposes has been chosen for the pixel module of IBL considering that three different sensor technologies (planar-silicon, 3D-silicon and diamond) are currently under investigation..\\ Two different format layouts were foreseen for different sensor types and they are shown in Fig.\ref{Fig:ModuleLayout}. Fig.\ref{Fig:ModuleLayout}-a shows the single-chip assembly (mainly for 3D and diamond) which presents a very narrow sensor edge, while the two-chip assembly  (for planar) with slightly greater dead edge is shown in Fig.\ref{Fig:ModuleLayout}-b.\\ For single-chip (two-chip) assemblies the nominal active coverage for particles normal to the beam is 98.8\% (97.4\%). An air gap of 100 $\mu$m (200 $\mu$m) between single-chip (two-chip) assemblies has been assumed to take into account the need for higher bias voltages for two-chip (planar-silicon) sensors. A 200 $\mu$m gap is considered sufficient to place polyimide insulation film if necessary, to electrically isolate adjacent modules.\\
Two n-in-n planar designs with different guard ring structures are considered. One of them has 450 $\mu$m wide guard ring structure regarded as conservative design. The other one has  a more aggressive slim edge with only 100 $\mu$m wide inactive area by shifting the guard ring structure under the active pixel area. Also a thin n-in-p planar design which utilizes the advantage of thinned sensors at given maximum bias voltage has been evaluated.\\
For 3D-silicon sensors the preferred option is the so called Passing-Through Active Edge \cite{Full3D} but also the Double-Type-Column approach \cite{FBK},\cite{CNM} has been taken into account.\\
For diamond sensors, the polycrystalline Chemical Vapor Deposition (pCVD) \cite{dPIX} technology has been chosen.\\
Looking to the sensors choice we have to consider that: 
a) Planar sensors require the lowest temperature and high bias voltage, but they have very well understood manufacturing sources, mechanical properties, relatively low cost, and high yield;
b) 3D sensors require the lowest bias voltage, intermediate operating temperature, and achieve the highest geometrical acceptance due to active edges, but their manufacturability with high yield and good uniformity must be demonstrated;
c) Diamonds require the least cooling, have low leakage current and capacitance and  have similar bias voltage requirements as planar sensors, but their manufacturability with high yield, moderate cost, and good uniformity must be demonstrated.\\
A significant factor in making the sensor choice will be the electrical performance with the FE-I4 front-end chip. The minimal stable operation threshold will be considered to be the parameter that determines the required charge at the end of the life  and not the single pixel white noise.\\
First prototypes available for testing are foreseen for autumn 2010 and irradiation and beam tests are scheduled for module's qualification during winter 2010-11.
\section{SUMMARY}
The Insertable B-Layer (IBL) will be the fourth low-mass pixel layer of the present ATLAS Pixel detector. The sensor technology will be chosen out of 3 different candidates: planar-Si, 3D-Si and diamond. First modules available for testing will be ready towards the end of 2010 and lab-measurements, irradiation and beam tests are foreseen for their qualification.\\ The scheduled installation for the detector is foreseen for end of 2014 and until then all IBL components need to be produced and tested individually and all together.

\end{document}